\documentstyle[12pt]{article}

\newcommand{\nc}{\newcommand}
\nc{\beq}{\begin{equation}}
\nc{\eeq}{\end{equation}}
\nc{\beqa}{\begin{eqnarray}}
\nc{\eeqa}{\end{eqnarray}}

\input epsf
\newwrite\ffile\global\newcount\figno \global\figno=1

\def\writedef#1{}

\def\figin{\epsfcheck\figin}\def\figins{\epsfcheck\figins}
\def\epsfcheck{\ifx\epsfbox\UnDeFiNeD
\message{(NO epsf.tex, FIGURES WILL BE IGNORED)}
\gdef\figin##1{\vskip2in}\gdef\figins##1{\hskip.5in}
\else\message{(FIGURES WILL BE INCLUDED)}%
\gdef\figin##1{##1}\gdef\figins##1{##1}\fi}

\def\figinsert{}
\def\ifig#1#2#3{\xdef#1{fig.~\the\figno}
\writedef{#1\leftbracket fig.\noexpand~\the\figno}%
\figinsert\figin{\centerline{#3}}\medskip\centerline{\vbox{\baselineskip12pt
\advance\hsize by -1truein\center\footnotesize{  Fig.~\the\figno.} #2}}
\bigskip\endinsert\global\advance\figno by1}
\def\endinsert{}

\begin{document}

\title{\large{\bf Gaugino Condensate and \\
Veneziano-Yankielowicz Effective Lagrangian}}

\author{Myck Schwetz\thanks{myckola@baobab.rutgers.edu}\\
{\small{Department of Physics and Astronomy, Rutgers University}}\\ 
{\small{Piscataway, NJ 08855-0849, USA}}\\
\\and \\\\
Maxim Zabzine\thanks{zabzin@vanosf.physto.se}\\
{\small{Institute of Theoretical Physics, University of Stockholm,}}\\
{\small{ Box 6730, S-11385 Stockholm, Sweden}}\\}

\date{}

\maketitle

\begin{picture}(0,0)(0,0)
\put(390,380){RU-97-83}
\put(390,365){USITP-97-14}
\end{picture}
\vspace{-24pt}

\begin{abstract}
We study the supersymmetric pure Yang-Mills theory
with semisimple Lie groups. We show that the general form 
of the gluino condensate is  determined solely by the
symmetries of the theory and it is in  disagreement with
the recently proposed existence of a  conformal phase 
in SYM theory. 
We discuss the peculiarities of
the Veneziano-Yankielowicz effective Lagrangian 
approach and explain  how it is 
related to the calculation of the gluino condensate. 
\end{abstract}

\newpage
\section{Introduction}

The exact solutions \cite{seiberg1, seiberg2, seiberg3}
for the IR Wilsonian effective theory of N=1 supersymmetric QCD (SQCD) 
are striking in their  self-consistency and seem to refuse any doubts 
of their validity. They brought about  a new direction of  model constructions
in the high energy particle phenomenology based on the notion of  
dynamical supersymmetry breaking. The nonzero gaugino condensates of 
the corresponding  supersymmetric Yang-Mills theory (SYM) play an important 
role in any  such a  model. 

Recently it was suggested  that the SYM theory
contains extra, chirally symmetric, vacuum  with zero gaugino condensate
\cite{kovshi}.
The argument was based upon an analysis of the Veneziano-Yankielowicz
effective Lagrangian for the SYM theory. 
If this phase indeed exists, then it has  drastic consequencies for the 
vacuum structure of SQCD. In particular, in the case of   $SU(N)$ SQCD 
with $N_f$ ($N_f \leq N$) flavours of massless matter fields there will 
be an extra disjoint point (at the origin) in the quantum moduli space of 
vacua. This will imply that there is no dynamical supersymmetry breaking 
in a large set of models which are believed to have it.

In this letter  this problem is studied from the point of view of the 
symmetries of the SYM theory. We derive the general form for a gaugino 
condensate and find that the existence of a chirally symmetric phase is not 
compatible with the symmetries of the theory. Also we discuss the applicability
of the Veneziano-Yankielowicz Lagrangian to the description of the low
energy SYM theories.

When this work was completed, a preprint by C. Csaki and H. Murayama 
\cite{CM} appeared which reaches some of the conclusions of this letter.

\section{SUSY Yang-Mills}

The SYM Lagrangian describing the gluodynamics of gluons $A_\mu$ and 
gluinos $\lambda_\alpha$ with general compact gauge group has the form 
\begin{equation}
\label{SYM}
{\cal L} ~=~ -\frac{1}{4g_0^2} 
G_{\mu\nu}^a G_{\mu\nu}^a
~+~ \frac{i}{g_0^2}\lambda_{\dot\alpha}^\dagger 
D^{\dot\alpha\beta}\lambda_\beta
~+~ \frac{i\theta}{32\pi^2} G_{\mu\nu}^a \tilde{G}_{\mu\nu}^a. 
\end{equation}
where $G_{\mu\nu}^a$ is the gluon field strength tensor, $\tilde{G}_{\mu\nu}^a$
is the dual tensor and $D^{\dot\alpha \beta}$ is the covariant
derivative and all quantities are defined with respect to the adjoint 
representation of the gauge group. This  Lagrangian may be written
in terms of  the  gauge superfield $W_\alpha$
with physical components ($\lambda_\alpha$, $A_\mu$) as follows
\beq
\label{SSYM}
{\cal L} ~=~ \frac{1}{8\pi} Im \int d^2\theta~\tau_0 W^{\alpha}W_{\alpha}~, 
\eeq
where the bare gauge coupling $\tau_0$ is defined to be
$\tau_0 = \frac{4\pi i}{g^2_0} + \frac{\theta_0}{2\pi}$.
The model possesses a discrete global $Z_{2C_2}$ symmetry\footnote{$C_2$
denotes the quadratic Casimir with $C_2 = 1$ normalization for the fundamental 
and anti-fundamental representations.}, a residual 
non-anomalous subgroup of the anomalous chiral $U(1)$. One way to see these
is to note that a chiral rotation of the gluino field becomes a symmetry if we 
combine it with a shift of the $\theta$ parameter 
\beq
\label{shift}
\theta \rightarrow \theta~ +~ 2C_2 \alpha,\,\,\,\,\,\,\,\,\,
\tau \rightarrow \tau ~+~ \frac{C_2}{\pi}\alpha 
\eeq
where $\alpha$ is a parameter of a chiral rotation. The physics of 
Yang-Mills theory is periodic in $\theta$ with period $2\pi$, no
compensation in $\theta$ is necessary if $\alpha = k\pi/C_2$ where $k$ is
an integer.

We will calculate the gluino condensate starting from the definition:
\beq
\label{def}
\langle \lambda^{a\alpha} \lambda^a_\alpha \rangle ~=~ \frac{\int
 DA\,D\lambda\,\,\lambda^{a\alpha}\lambda^a_\alpha\,e^{-\int {\cal L}
d^4 x}}{\int DA\,D\lambda\,\,e^{-\int {\cal L} d^4 x}}.
\eeq
One may introduce an external  current $J$ and rewrite (\ref{def})  as follows
\beq
\label{defcur}
\langle \lambda \lambda \rangle~ =~ \frac{\delta}
{\delta J} Z[J]\,\Big|_{J=0}~ =~ \frac{\delta}{\delta J}\, [ \log
 \int DA\, D\lambda\,\,e^{-\int {\cal L}d^4 x + \int \lambda\lambda J
\, d^4 x}]\,\Big|_{J=0}~.
\eeq
Using the symmetries of the theory we will examine $Z[J]$. Let us discuss
 how one should interpret the generating functional $Z[J]$. 
Suppose that $Z[J]$ is regularized in such a way that  the discrete
 symmetry $Z_{2C_2}$ are maintained. 
Then the generating functional 
depends on the bare coupling constant $\tau_0$, the cut-off M (the 
parameter of regularization) as well as the current. The discrete 
symmetry is realized as follows
\beq
\label{symm}
Z[\tau_0 + \frac{C_2}{\pi} \alpha,\,M,\,e^{-2i\alpha}J] ~=~ Z[\tau_0,\,M,\,J]~,
~~~~~ \alpha~=~\frac{k\pi}{C_2}~, ~~k\in Z
\eeq
where $\alpha$ is a parameter of  the chiral rotation
\beq
\label{rotation}
\lambda \rightarrow e^{i\alpha} \lambda~.
\eeq
Let us take the derivative with respect to $\alpha$ on the left hand side  
 of (\ref{symm}) and set $\alpha$  equal to zero afterwards
\beq
\label{eq}
\frac{dZ}{d\alpha}\,\Big|_{\alpha=0} ~=~0~~~ \Longleftrightarrow
~~~\frac{dZ}{d\tau_0}~=~ \frac{2\pi i}{C_2} \int
 \langle \lambda \lambda \rangle_J\,J\,d^4 x
\eeq
where $\langle \lambda \lambda \rangle_J$ is defined by the 
 following expression
\beq
\label{expression}
\langle \lambda \lambda \rangle_J ~=~ \frac{\int DA\,D\lambda \,\,
\lambda\lambda\,\,e^{-\int {\cal L} d^4 x + \int \lambda \lambda J
\,d^4 x}}{\int DA\,D\lambda\,\,e^{-\int {\cal L} d^4 x + \int \lambda
 \lambda J\,d^4 x}}~,\,\,\,\,\,\,J\neq 0~.
\eeq
The differential equation (\ref{eq})  cannot  be solved directly for the 
generating functional because  it is  unknown 
how to formulate the initial-value problem (the Cauchy problem)
for the functional integral. Let us
 take the functional derivative $\delta/\delta J$ of (\ref{eq}) at $J=0$.
This results in
\beq
\label{eq2}
\frac{d}{d\tau_0}\, \langle \lambda \lambda \rangle~ =~ \frac{2\pi i}
{C_2}\, \langle \lambda \lambda \rangle.
\eeq
Using a dimensional argument, one can write down the general
solution of  (\ref{eq2})
\beq
\label{solution}
\langle \lambda^{a\alpha}(x) \lambda^a_\alpha(x) \rangle ~=~
c\, M^3 f(Mx)\, e^{\frac{2\pi i}{C_2} \tau_0}~,
\eeq
where $c$ is the constant of  integration and $f$ is an 
arbitrary function of the dimensionless combination $Mx$. 
Supersymmetry requires  the gluino condensate to be independent
of the coordinate x, i.e. to be constant \cite{NSVZ}. 
Thus the discrete symmetry $Z_{2C_2}$ and the 
supersymmetry determine the general form of the gluino condensate
 to be
\beq
\label{form}
\langle \lambda \lambda \rangle ~=~c\, M^3 \,
e^{-\frac{8\pi^2}{C_2 g_0^2}}\,e^{\frac{i\theta}{C_2}}~.
\eeq
Recall that the chiral rotation (\ref{rotation}) and the 
shift in $\theta$ (\ref{shift}) leave the 
theory invariant if $\alpha=k\pi/C_2$\,, $k\in Z$. At $\theta = 0$ the 
general form of the gluino condensate is 
\beq
\label{form1}
\langle \lambda \lambda \rangle ~=~ c\, M^3 e^{-\frac{8\pi^2}
{C_2 g_0^2}}\, e^{\frac{2\pi i k}{C_2}}~,\,\,\,\,k\in Z~.
\eeq
The gluino condensate does not acquire an anomalous 
 dimension under renormalization since $\langle \lambda
 \lambda \rangle$ is the lowest component of the superfield $W^2$ 
and the upper component of the same superfield contains the trace of
the stress tensor. Thus the gluino 
 condensate needs no subtractions and the renormalized 
 expression for the gluino condensate is
\beq
\label{renorm}
\langle \lambda \lambda \rangle~=~ c\, M^3 \, e^{-\frac{8\pi^2}
{C_2 g^2}}\, e^{\frac{2\pi i k}{C_2}}~,
\eeq
where now $M$ is the mass scale (the point of renormalization) and
 $g$ is the renormalized coupling constant (the running coupling constant).

It is convenient to define the strong coupling scale $\Lambda$
 at which the expression for the one-loop effective 
 running coupling constant diverges\footnote{In any particular
 perturbative scheme a scheme-dependent constant can be added
 to the one-loop expression for the running coupling
 constant. This constant generates a constant rescaling of
 $\Lambda$ and can be ignored, since it can be absorbed by 
 a redefinition of $M$.}
\beq
\label{strong}
\Lambda^{b_0}~=~ M^{b_0} \,e^{-\frac{8\pi^2}{g_0^2}}
\eeq
where $b_0$ is the  coefficient in the one-loop 
 $\beta$-function ($b_0 = 3C_2$). In terms of the strong
 coupling scale the gluino condensate has the form
\beq
\label{form3}
\langle \lambda \lambda \rangle ~=~ c\, \Lambda^3\,
e^{\frac{2\pi i k}{C_2}},\,\,\,\,\,k\in Z~.
\eeq
If the constant $c$ is not zero then we can conclude that
 the SYM theory has $C_2$ different
values of the gluino condensate. This statement is just
 a consequence of the global symmetries. 

Alternatively we may derive the general form of the gluino 
condensate using the  idea that the coupling constant
$\tau_0$  can be treated as a spurion chiral superfield. 
Let us introduce the generating functional as follows
\beq
\label{gener}
\tilde{Z}[\Phi,\,\bar\Phi,\,M]~=~ \log \int DV \,\,\,
e^{-\frac{1}{8\pi}\int d^4 x\,Im\,(\int d^2 \theta\,\Phi WW)}~,
\eeq
where $M$ is the parameter of the regularization and 
 $\Phi$ is an external chiral field
\beq
\label{chiral}
\Phi ~=~ \phi ~+~ \theta^\alpha \Psi_\alpha ~+~ \theta\theta F~.
\eeq
It is important to note that $\tilde Z$ is not a holomorphic 
 function of $\Phi$ because in the definition of SYM 
 Lagrangian we use the imaginary part. In terms of this generating
functional the gluino condensate has the form
\beq
\label{functional}
\langle \lambda \lambda \rangle ~=~ \frac{\delta}{\delta F}
\tilde Z [\Phi,\,\bar\Phi,\,M]\,\Big|_{\begin{tiny}
                                \begin{array}{lrc}
                                F=\bar{F}=0\\
                                \Psi=\bar\Psi=0\\
                                \phi=\tau_0, \bar\phi=0
                                \end{array}
                                \end{tiny}}~.
\eeq
Using the symmetries of the theory we can construct the 
 generating functional $\tilde Z$. The discrete symmetry
 $Z_{2C_2}$ is realized in the following way
\beq
\label{symetries}
\tilde Z [\Phi+\frac{C_2}{\pi}\alpha,\,\bar\Phi+\frac{C_2}
{\pi}\alpha,\,M]~ =~ \tilde Z [\Phi,\,\bar\Phi,\,M]~,\,\,\,
\alpha~=~\frac{k\pi}{C_2}~,\,\,k\in Z~.
\eeq
The supersymmetry is manifest in this superfield formulation.
 One can write the Taylor expansion for the generating 
 functional in superspace
\beq
\label{Taylor}
\tilde Z [\Phi,\,\bar\Phi,\,M] ~=~ \int d^4 x\,d^2 \theta \,G_1(\Phi)
~+~ \int d^4 x\, d^2 \bar\theta\,G_2(\bar\Phi)~ +~ \int d^4 x\, d^2\theta\,
d^2 \bar\theta \, G_3(\Phi,\,\bar\Phi)~,
\eeq
where $G_1$, $G_2$ and $G_3$ are some functions which can be
 fixed by the discrete symmetry and dimensional arguments
\beq
\label{functions}
G_1(\Phi)~=~A M^3 e^{\frac{2\pi i}{C_2} \Phi}~,\,\,\,\,
G_2(\bar\Phi)~=~ B M^3 e^{-\frac{2\pi i}{C_2}\bar\Phi}~,\,\,\,\,
G_3(\Phi,\,\bar\Phi)~=~ M^2 f(\Phi-\bar\Phi)~,
\eeq
where $A$ and $B$ are some constants and $f$ is an arbitrary 
function. Inserting (\ref{Taylor}) and (\ref{functions}) in expression 
 (\ref{functional}) we can imediatly obtain the general form of the 
gluino condensate
\beq
\label{form4}
\langle \lambda \lambda \rangle ~=~\frac{2\pi i}{C_2} A \, M^3 
e^{\frac{2\pi i}{C_2}\tau_0}
\eeq
As we see, in both approaches the general form of the gluino
condensate is fixed by only two tools, supersymmetry
and  discrete  global symmetry. 
   
The equation (\ref{eq2}) has two solution: 
(\ref{form3})  with $c\neq 0$ (or $A \neq 0$ in (\ref{form4}) ) and the 
trivial $\langle\lambda\lambda\rangle =0$ ($A = 0$). 
Can one further eliminate possibilities?
Let us look closely at the
phase which is described by the zero value of the gluino condensate.
There are no mass parameters in this phase and it should
be conformal or more precisely, superconformal, since we 
assume that supersymmetry is unbroken. 
To see this consider the dilation transformation $ x'= e^{-t} x\, ,
\theta' = e^{-t/2} \theta \, , \bar\theta' = e^{-t/2} \theta$.
Classically it is a symmetry of the modified SYM Lagrangian in 
(\ref{gener}) but quantum mechanically - not. The path integral measure 
in (\ref{gener}) is not invariant under such a dilation:
\beq
\label{Vdil}
DV(x',\theta',\bar \theta')~=~DV(x,\theta,\bar \theta)\,
exp\left({{3C_2 \over 4\pi} t \,\int d^4 x\, d^2 \theta\, W^2}
~+~ h.c. ~+~ O(1/M^4)\right)~.
\eeq
where $O(1/M^4)$ refers to higher dimensional $D$-terms supressed by
 powers of the cut-off. Therefore the $F$-term is exact and effectively
one may consider the dilaton transformation as an imaginary shift in 
 the spurious superfield $\Phi$: $\Phi \rightarrow \Phi ~+~
i{3C_2 \over 2\pi} t$ and correspondently the chiral transformation as a 
 real shift in the spurious superfield $\Phi$: $\Phi \rightarrow \Phi ~+~
{C_2 \over \pi}\alpha$. If we consider the phase with zero gluino
 condensate then the $F$-terms will disappear in generating functional 
 (\ref{Taylor}). This functional will be invariant under the full chiral
 transformation $U(1)$. 
Since the dilaton transformation is in the same supermultiplet with 
R-transformation the $D$-term which is not invariant under dilations
has to be zero in this phase too.
 We see that phase with zero condensate has to be superconformal, if 
 it exists.

In other words, the superfield $\Phi$ is the source for the anomaly 
 multiplet $W^2$. The generating functional (\ref{Taylor}) is the
 functional for the connected Green functions of the composite 
 superfiled $W^2$. So the Legendre transform of (\ref{Taylor}) will
 give the effective action which satisfies the anomalous chiral
 Ward identities. In case when $F$-terms are absent in
 (\ref{Taylor}), the effective
 action is identicaly zero. The absense of anomalies\footnote{The
 zero effective action for the multiplet of anomalies one can interpret
 only as absence of anomalies.} contradicts
 with the defining properties of the pure SYM theory. There is no such 
kind of IR fixed point when anomalies disappear like in some SQCDs.  
The proposal by Kovner and Shifman \cite{kovshi} disagrees with the global 
symmetries of the SYM theory.

At the end, we want to discuss the properties of R-currents in pure SYM
 and in conformal window of $SU(N_c)$ SQCD. 
We established that the phase with $\langle\lambda\lambda\rangle =0$
should necessarily be superconformal. At such a superconformal fixed point 
the axial $R$-current 
($J_{\alpha\dot\alpha}={2 \over g^2} 
Tr(\lambda_\alpha \bar\lambda_{\dot\alpha})$)
must be conserved. Therefore in pure SYM theory there is
no such kind of the fixed point as $R$-symmetry anomaly matching 
conditions will not be satisfied. 
In general not all solutions to a differential equation 
(whether this is the equation (\ref{eq2}) or minimization equation on the 
effective potential considered in the next section) are physical. 
The physical ones should satisfy initial or boundary and all other types 
of conditions
imposed on the model. And anomaly matching is the sieve in the 
case. It is interesting to note that in the conformal window of
$SU(N_c)$ SQCD the divergence of the axial current $J_{\alpha\dot\alpha}$
from the multiplet of anomalies becomes zero too. But unlike in SYM, 
there is a conserved anomaly-free linear combination (with coupling 
dependent coefficients) of anomalous currents. When the theory flows 
to the fixed RG point the anomaly-free combination becomes pure axial 
current \cite{kogshiv}. Thus there is no new global symmetry appearing
at a fixed point. 

\section{VY Lagrangian}

We believe that in SYM theory only colorless asymptotic states  exist 
and that a mass gap is dynamically generated. So at low energy one has to describe 
the theory using the new degrees of freedom. The gluons and gluinos must 
disappear from the low energy description. It is believed that the relevant 
composite degree of freedom can be  naturally constructed \cite{VY} in terms of 
the chiral superfield 
\begin{equation}
S ~=~ \frac{3}{32\pi^2}W^2 \equiv ~ \frac{3}{32\pi^2}\,\mbox{Tr}\,W^2~,
\end{equation}
where the color trace above is in the adjoint representation. Using the path 
integral formalism one can write
\beq
\label{pathint}
W(J_{S}, J_{\bar{S}}) ~=~ \log \int DV e^{-\int d^4 x\, {\cal L} + \int
d^2 \theta\,d^4 x\, SJ_{S} + \int d^2\bar\theta\,d^4 x\, J_{\bar S}\bar S}~,
\eeq
where $W(J_{S}, J_{\bar S})$ is the generating functional for the connected 
Green function of the composite operators. Here, $J_{S}$ and $J_{\bar S}$ are
the chiral and antichiral currents
\beq
\label{currents}
J_{S}~=~J_1(x^+) ~+~ \theta^{\alpha}J_{2\alpha}(x^+)~+~\theta\theta J_3~(x^+),
\eeq
\beq
\label{currents2}
J_{\bar S}~=~ J_1^*(x^-) ~+~ \bar\theta^{\dot\alpha} \bar{J}_{2\dot\alpha}(x^-)
~+~ \bar\theta \bar\theta J_3^*~(x^-),
\eeq
where we use the standard notation $x^+ = x+i/2\,\theta\bar\theta$ and 
$x^- = x - i/2\,\theta\bar\theta$.
The effective action $\Gamma(S,\bar S)$ is defined in standard way  as a 
 Legendre transform of $W(J_{S},J_{\bar S})$
\beq
\label{transform}
\Gamma(S, \bar S) ~=~ W(J_S, J_{\bar S}) ~-~ \int d^2\theta\,d^4 x\,J_{S}S
~-~ \int d^2\bar\theta\,d^4 x\,J_{\bar S}\bar S.
\eeq
One can consider the Taylor expansion of the effective action in superspace
\beq
\label{Taylor2}
\Gamma(S, \bar S) ~=~ \int d^2 \theta\, d^2 \bar\theta\, F_1(S,\bar S) ~+~
\int d^2 \theta F_2(S)~+~ \int d^2 \bar\theta F_3(\bar S)
\eeq
and fix the functions $F_1$, $F_2$ and $F_3$ using the chiral Ward \cite{shore} 
identities
\beq
\label{func1}
F_1(S, \bar S) ~=~ (S\bar S)^{1/3} f\left(\frac{S^{1/3}}{(\bar D^2 \bar S^{1/3})^{1/2}},
\, \frac{\bar S^{1/3}}{(D^2 S^{1/3})^{1/2}}\right) ~,
\eeq
\beq
\label{func2}
F_2(S)~=~\frac{C_2}{3}\left(S\, ln(\frac{S}{\Lambda^3}) ~-~ S \right)~,
\eeq
\beq
\label{func3}
F_3(\bar S)~=~\frac{C_2}{3}\left(\bar S\, ln(\frac{\bar S}{\Lambda^3})~-~\bar S
\right)~,
\eeq
where $f(x,y)$ is an arbitrary function two variables , satisfying 
the reality condition
\beq
\label{cond}
f^* (x, y) ~=~ f(y,x)~.
\eeq
This is the most general solution of the problem. Starting from it 
one can immediately write the zero momentum form of the effective low energy 
Lagrangian (Veneziano-Yankielowicz Lagrangian) \cite{VY}
\begin{equation}
\label{VYL}
{\cal L} ~=~\frac{9}{a}\left( \bar S S \right)^{1/3}\Big|_D ~+~ 
\frac{C_2}{3} \left(S\ln (S/ \Lambda^3~-~S) \right)\Big|_F+ \, \mbox{h.c.}~,
\end{equation}
where 
$a$ is a numerical parameter related to the wave-function renormalization
$Z(p^2)$ taken at zero momentum.

Let us discuss the relation between the Veneziano-Yankelowicz action and 
the calculation of the gluino condensate. The chiral field $S$ can be 
written in the standard form
\beq
\label{chiral1}
S~=~\phi(x^+)~+~\theta^\alpha \psi_\alpha(x^+) ~+~ \theta\theta F(x^+)
\eeq
and be inserted in (\ref{VYL}). The interesting part of the Lagrangian
 including the $F$-field is
\beq
\label{part}
V(\phi,\phi^*, F, F^*) =-\frac{a}{9}(\phi \phi^*)^{-2/3} F F^* ~-~ 
\frac{C_2}{3}F\, ln\frac{\phi}{\Lambda^3} ~-~
\frac{C_2}{3}F^*\, ln \frac{\phi^*}{\Lambda^3} ,
\eeq 
where $F$ is regarded as an auxiliary field. Elimination of the
 auxilary field leads to the original VY scalar potential
\beq
\label{potential1}
V(\phi, \phi^*) = \frac{1}{a} |\phi|^{4/3} \left(ln^2 \frac{|\phi|^{C_2}}
 {\Lambda^{3C_2}} + C_2^2 (\arg \phi)^2\right) \geq 0,
\eeq
which has the $C_2$ standard minima and one exotic at $\phi=0$.
 This answer is natural, because we reproduce all solution of the 
differential equation (\ref{eq2}). The differential equation (\ref{eq2})
 is just a consequence of the global symmetries of SYM. The potential 
 (\ref{potential1}) comes as a solution of Ward identities. As it was 
pointed out one has to look carefully at all solutions before
 identifying the true vacua of the theory. 
Note that all derivatives of the potential $V(\phi, \phi^*)$ at 
$\phi=\phi^*=0$ are not well defined. The potential $V(\phi, \phi^*)$ 
was obtained as a solution of Ward identities which are the 
functional-differential equations. If the field $\phi$ is regarded as a
 constant field then the Ward identities are
 just the differential equations. Since the derivatives are not well defined, 
 the potential is not a solution of Ward identities at the origin. 
So the solution $\phi=0$ is not compatible with Ward identities. 

The non-constant function $f$ 
in the action (\ref{Taylor2})-(\ref{func3})
 leads to the propogation of the auxilary field $F$. As was noted 
 in \cite{shore} the effective potential will include $\phi$
 and $F$-fields in this case and it will not be
 bounded from below. One must eliminate the auxilary field\footnote
 {In the case of general function $f$, it is not possible to solve
 explicitly equations of motion for auxilary fields. If lucky,
one may utilize perturbation theory expanding in powers of 
some small parameter and higher derivative terms.}. After
 their elimination they lead to a theory with nonlocal interactions 
since the auxilary fields have nontrivial dynamical equations. 
 But if  we are interested mainly in the effective potential (the 
 effective action with its external momemeta set to zero), we neglect
all higher order derivative terms.
  
We would like to finish with a few remarks about the effective Lagrangian
approach. Historically, many of the non-perturbative properties was discovered
by effective Lagrangian approach which is based on the construction
of the most general possible form of the action satisfying the 
non-anomalous and anomalous Ward identities of the underlying theory. 
 From this general form of the effective action one can extract, 
for example,  the possible form of condensates. We have to remember that 
the notion of the effective Lagrangian is defined correctly\footnote{By 
this we mean that one keeps all usual properties of the QFT like a cluster 
decomposition, a spectral representation and etc (the Wightman QFT). 
The Legendre transform of the effective Lagrangian provides
the generating functional for the connected Green functions of the relevant 
fundamental or composite fields.} when we ``sit'' 
in one of the vacua. In this sense the effective Lagrangian approach study 
a set of different Lagrangians, because the VY Lagrangian is not a 
singlevalued function of the fields and there no explicit $Z_{2C_2}$ symmetry.
These features were pointed out in \cite{kovshi} as
defects in the VY Lagrangian. These problems automatically can be fixed if 
we consider the effective Lagrangian in one of $C_2$ vacua.  
The possible physical  information can still be  extracted from the VY
effective Lagrangian, in particular some mass ratios of the bound states
can be predicted \cite{EHS}. In the end, this approach could be tested on 
the lattice.

\section{Summary}

We have derived the gaugino condensates for SYM theories starting from the
constraints imposed by supersymmetry and other global symmetries.
The result completely agrees with the exact result in SQCD.
We have argued that the vacuum with zero gaugino condensate corresponding 
to the possible conformal phase of SYM theory is not compatible with the
global symmetries of the theory.  Therefore the exact results in SQCD
are not modified and phenomenological models based on 
 dynamical supersymmetry breaking remain viable. 
Finally we commented on the  applicability of the Veneziano-Yankielowicz 
effective Lagrangian approach.

\begin{flushleft} {\Large\bf Acknowledgments} \end{flushleft}

\noindent The authors are grateful to Nick Evans, Stephen Hsu,
Ulf Lindstr\"om, Nathan Seiberg and 
Graham Shore 
for useful discussions and comments.
The authors thank Isaac Newton Institute for the kind hospitality while
some of this work was carried out. 
This work was in part funded by the NSF grant number NSF-PHY-94-23002 and
by the grant of the Royal Swedish Academy of Sciences.


\end{document}